# DYNAMIC CONDENSATION OF WATER AT CRACK TIPS IN FUSED SILICA GLASS

M. Ciccotti[1,*], M. George[1], V. Ranieri[1], L. Wondraczek[2] and C. Marlière[3]

[1]Laboratoire des Colloïdes, Verres et Nanomatériaux, Université Montpellier 2, CNRS, France
[2]Corning European Technology Center, Corning SAS, Avon, France
[3]Géosciences Montpellier, Université Montpellier 2, CNRS, France

**ABSTRACT**

Water molecules play a fundamental role in the physics of slow crack propagation in glasses. It is commonly understood that, during stress-corrosion, water molecules that move in the crack cavity effectively reduce the bond strength at the strained crack tip and, thus, support crack propagation. Yet the details of the environmental condition at the crack tip in moist air are not well determined. In a previous work, we reported direct evidence of the presence of a 100 nm long liquid condensate at the crack tip in fused silica glass during very slow crack propagation ($10^{-9}$ to $10^{-10}$ m/s). These observations are based on *in-situ* AFM phase imaging techniques applied on DCDC glass specimens in controlled atmosphere. Here, we discuss the physical origin of the AFM phase contrast between the liquid condensate and the glass surface in relation to tip-sample adhesion induced by capillary bridges. We then report new experimental data on the water condensation length increase with relative humidity in the atmosphere. The measured condensation lengths were much larger than what predicted using the Kelvin equation and expected geometry of the crack tip.

## 1. INTRODUCTION

The brittleness of glass is the most limiting factor in its widespread applications. Long-term behavior of vitreous materials is particularly important with respect to their use in architectural elements and their role in the encapsulation of nuclear wastes. However, despite of several decades of extensive research, most of the knowledge of sub-critical crack propagation in glasses is still based on a phenomenological ground, and dates back to the 1970s.

It is commonly understood that water molecules play the fundamental role in slow crack propagation in glasses. The dissociative adsorption of water molecules on strained siloxane bridges at the crack tip is the basic ingredient of the stress-corrosion theory for sub-critical crack growth [1]. The functional dependence of crack velocity on partial water vapor pressure was determined by Wiederhorn [2,3], along with the presence of a transport limited region in the $K_I$-v curve. His model is able to reproduce a wide range of experimental data concerning sub-critical fracture in glass, and is still the basis of most technological considerations. His observations could be, in principle, interpreted by different approaches based on stress-enhanced thermal activation [4], combined with related transport or diffusion-limited processes. On the other hand, Tomozawa [5] and others explained water influence on fracture propagation as a local weakening effect due to stress-enhanced water diffusion into the strained neighborhood of the crack tip. Thus, the fundamental rupture mechanism which is generally acting at the tip is still debated, and the influence of glass chemical composition on the fracture properties presently far from understood [4,6,7].

The reason for this lack of understanding lies in the high homogeneity of the glass structure down to sub-micrometric scales. Physical mechanisms of fracture have to be considered at the nanometer scale, where they are heavily mixed with chemical processes.

Development of scanning probe microscopes provided new insights during the last years and decades by allowing the glass surfaces to be probed down to nanometric spatial resolution. In combining atomic force microscopy (AFM) with low-speed fracture experiments, the nanometric neighborhood of the crack tip in glass samples can be observed during crack propagation in a carefully controlled atmosphere [8]. When the crack propagation velocity decreases below $10^{-9}$ m/s (a typical limit for optical techniques) AFM studies of the vicinity of a crack tip (at its intersection with the external surface of the sample) have revealed distinct phenomena that are believed to play a major role in the comprehension of the physics of slow fracture [8]. E.g., the hypothesis of nucleation, growth and coalescence of nanometric cavities as the origin of slow cracks in glasses is based on such observations [9]. The importance of stress-enhanced diffusion of mobile alkali ions was put into evidence by direct observation of a micrometric parabolic front of sodium nodules accompanying the crack tip propagation in soda-lime glasses [10].

More recently, the occurrence of liquid condensates at the crack tip was observed *in-situ* in silica glasses [11] by AFM phase imaging. The present work will first discuss the physical origin of the observed phase contrast by relating it to changes in the local adhesion between the AFM tip and the glass surface caused by the presence of the liquid condensate. New experimental data on the conditions of formation of the condensate and the dependence of its size on relative humidity will then be presented. Complementary observations that can be found in literature will be considered in this context.

## 2. EXPERIMENTAL PROCEDURE

The experimental setup is illustrated in Fig. 1. The DCDC [12] technique was employed to create mode I tensile cracks with highly controllable speed in silica glasses. For an applied stress $\sigma$, the stress intensity factor is given by $K_I = \sigma a^{1/2}/(0.375c/a+2)$, where $a$ is the radius of the hole and $c$ is the length of the crack [12]. Thus, the crack speed will actually decrease with increasing length, which makes this technique the best choice to fully control crack propagation down to a velocity of $10^{-12}$ m/s in a long-term experiment. The load cell is mounted on the two-dimensional high-precision positioning stage of an AFM (D3100, Veeco).

Parallelepipedic DCDC samples (4x4x40 mm³) of pure silica glass (Suprasil 311, Heraeus, Germany, bulk $OH^-$ content of 200 ppm) were polished with $CeO_2$ to a RMS roughness of 0.25 nm (for an area of 10x10µm²) and a hole of 1 mm in diameter was drilled at their center to trigger the start of the two symmetric fractures shown in Fig. 1. Generation and propagation of the crack were performed at 22.0±0.5 °C in a carefully controlled atmosphere of mixtures of pure nitrogen and pure water vapor.

The experiment was made in two steps: in the first one the $K_I$-v curve was measured at constant relative humidity (RH), while the crack velocity slowing down to around $10^{-9}$ m/s. Then a very slow drift in the relative humidity was induced to produce cycles between ~ 1 % and 80 % (± 2 %) during the four weeks duration of a typical experiment. The load was possibly adjusted to maintain the crack velocity between $10^{-10}$ and $10^{-9}$ m/s. AFM measurements were performed in a high amplitude resonant mode ('tapping' mode) at different magnifications ranging from 75x75 nm$^2$ to 5x5 µm$^2$.

## 3. AFM METHODS

In a precedent paper, we already described the combination of topographic and phase contrast images [11]. The observed phase contrast between the condensate region and the intact glass surface was attributed to changes in the adhesive interactions between the AFM tip and, respectively, the liquid condensate and the silica glass surface recovered by an adsorbed water layer the thickness of which is in nanometric range [13,14].

When operated in tapping mode (amplitude modulation), the actuator of the AFM cantilever is stimulated by a sinusoidal electric signal $E(t) = E_0 \cos(\omega t)$ with constant parameters: the excitation frequency $\omega$ is chosen close to the resonant frequency $\omega_0$ ($f_0 = \omega_0/2\pi \sim 330$ kHz); the excitation amplitude $E_0$ is set to obtain a free tip oscillation amplitude $A_0 \sim 30$ nm at resonance. During the scan of the sample's surface, a feedback loop acts on the vertical position of the whole cantilever in order to maintain the tip oscillation amplitude at a set point value $A_{sp} \sim 20$ nm. The related vertical displacement of the cantilever provides the vertical topographic signal.

Although the tip-sample interaction is non-linear, the motion of the tip can be well approximated by a harmonic oscillation $z(t) = A \cos(\omega t + \theta)$. The phase delay $\theta$ between the stimulation and the oscillation of the AFM cantilever (phase signal) can thus be related to the energy $E_{diss}$ which is dissipated due to the tip-sample interaction [15]:

$$\sin \vartheta = \frac{\omega}{\omega_0} \frac{A_{sp}}{A_0} + \frac{Q E_{diss}}{\pi k A_0 A_{sp}} \qquad (1)$$

where $Q \sim 600$ and $k \sim 40$ N/m are the quality factor and the stiffness, respectively, of the AFM cantilevers that were used for the experiment (*OLTESPA ESP Series Probes*, Veeco).

From phase contrast data reported in [11], only qualitative statements about the increase of the dissipated energy on the condensate region were derived. In order to obtain quantitative information (applying Eq. 1), an external lock-in preamplifier (Perkin-Elmer 7280) was connected to the AFM controller providing phase measurements with an uncertainty of 1°.

## 4. RESULTS

Comparing quantitative phase images with amplitude-phase-distance curves measured on the glass surface for different humidity values, the origin of the changes in the tip-surface interactions at elevated humidity (noted as "instabilities" in ref [11]) can be clarified and is now detailed.

The non-linear nature of the tip-sample interaction in 'tapping' mode may lead to the presence of multiple solutions for the amplitude and phase of the tip oscillation, depending on the cantilever drive parameters and on the details of the interaction [16, 17]. In particular, two main modes can be distinguished: (1) a lower amplitude mode, also called 'Non Contact' (NC) mode, in which the force acting on the AFM tip is prevalently attractive, and (2) a higher amplitude mode, also called 'Intermittent Contact' (IC) mode, in which the tip experiences a strong repulsive interaction with the sample at each vibrating cycle.

The NC (resp. IC) mode, has a typical phase delay $\theta < -90°$ (resp. $\theta > -90°$) and it is generally obtained for small (resp. large) free oscillation amplitude and a weakly (resp. heavily) reduced set point amplitude. The conditions of transition between these two modes can be very complicated and have been the subject of several studies [17,18]. A recent work from Zitzler [19] has shown that the critical oscillation amplitude that triggers the transition between IC and NC modes on a silica glass surface is an increasing function of the relative humidity and, thus, of the thickness of the condensed water layer, according to experimental investigations performed by Beaglehole and Christenson [20]. The estimation of the energy which is dissipated in the tip-surface interaction is consistent with a model based on the formation and rupture of a capillary neck between the AFM tip and the nanometric water layer. From this, the observed phase behavior will be interpreted now.

In Fig. 2 two typical phase images of the crack tip at low and high humidity values are shown. At 1% RH (resp. 70% RH), the phase signal on both the silica surface and the condensate region is higher (resp. lower) than -90°, the phase contrast on the condensate relative to silica surface being negative (resp. positive): the two images thus correspond respectively to IC and NC mode, but the dissipated energy is, in both cases, enhanced in the condensate region (cf. Eq. 1). Since the oscillation amplitude was kept constant during the measurements, the phase signal behavior on the silica surface is consistent with results from [19]: the IC mode (resp. NC mode) prevails for lower (resp. higher) RH values, or equivalently for thinner (resp. thicker) adsorbed water layer on the glass surface.

The presence of the capillary water condensation inside the crack is expected to induce a stronger energy dissipation due to the increased amount of available water. In Fig. 3 we report the evolution of the phase signal on both the silica surface and the condensate region during the experiment, along with the values of relative humidity and observed length of the condensate. The phase curves present three regimes: (1) from day 0 to day 8 (while humidity is gently rising from dry conditions) IC behavior is observed like in Fig. 3.a; (2) from day 8 to day 15 (including and following the high humidity stage) NC behavior like in Fig. 3.b is dominant; (3) from day 15 to day 18 (when the dry atmosphere is re-established), the AFM flips back to IC mode on the silica surface, but it remains in NC mode on the condensate region, indicating that the evaporation of the condensate is significantly slower than on the silica surface. The images in this regime are thus mixed mode images, where the mode transition is triggered by the condensate borders. The unstable behavior reported in [11] can be explained as a consequence of the hysteretic behavior of mode transitions [16].

Fig. 3c reports the evolution of the length of the capillary condensate. Due to uneven contrast in different modes and to the 10 nm radius of the AFM tip, the incertitude on the absolute condensate length can be estimated to be 20 nm. Yet, the condensate's length grows regularly with increasing relative humidity from 70 nm to 200 nm. When decreasing the relative humidity, the length rapidly decreases to 140 nm, then no further reduction is observed in the five following days. The presence of condensate at the crack tip can thus not be excluded, even in dry conditions, when a sample was exposed to humidity previously – a fact that has to be considered in crack propagation studies.

## 5. DISCUSSION

Capillary condensation of water at ambient humidity is expected to occur at the crack tip in silica glass when the local crack opening is smaller than twice the Kelvin radius [21]:

$$r_k = \frac{\gamma.V}{RT\log(P/P0)} \qquad (2)$$

where $\gamma$ is the surface energy of water, $V$ the molar volume of liquid water, $T$ the temperature and $P/P_0$ the relative humidity. At ambient conditions (21°C, RH around 45%, P=1atm), the Kelvin radius for water is of the order of a nanometer, which is also the typical value for the radius of curvature of the very sharp crack tips of glasses. The Kelvin radius is, however, increasing with increasing relative humidity. If in the present conditions (very low crack speed) equilibrium between condensate and surrounding atmosphere is assumed, the condensate length can be expected to grow with relative humidity until a distance from the crack tip where a crack opening equal to twice the Kelvin radius is achieved. A rough estimate based on a parabolic crack shape yields a maximum condensate length of 20 nm at 70% RH, a value that is significantly smaller than what is observed here.

However, the above estimation is too simple: several effects perturb the physics of condensation at such small scales and need to be taken into account. Firstly, the internal surfaces of the crack must be considered as wetted by a thin water layer, which causes the liquid condensate to grow to longer distances. The thickness of this layer may be large, but it is not easy to access. In the presence of negative charges on the fracture surfaces it can easily grow to several molecular layers [22]. Secondly, the effect of hydration forces and thirdly the presence of a small amount of soluble contaminants in the film are supposed to have distinct effects onto capillary condensation in such highly confined conditions [22]. Thus a complete understanding requires further studies: one parameter we plan to study too is the effect of the conditions of preparation of the free surface (role of grinding, etc…). Finally, roughness at the nanometer scale and below must be taken into account. Local changes in radii of curvature of the internal crack surfaces can translate into irregularities in the water film thickness, which could either provide local barriers to wetting, or induce the nucleation of capillary bridges in the crack. Bocquet *et al.* [23] have shown that the dynamics of formation of capillary bridges in humid granular materials are determined by the nanometric roughness of the particles. Here, too, further work is required to quantify this effect.

## 6. CONCLUSION

Application of a recently developed technique based on quantitative AFM phase imaging of the vicinity of a crack tip in glasses has permitted to demonstrate nanoscale-formation of hydrous capillary condensation inside the opening of a crack which propagates at velocities between $10^{-10}$ and $10^{-9}$ m/s. The condensate length was shown to increase from 50 to 200 nm for relative humidity rising from 1% to 70%. The condensate was further shown to persist several days in dry conditions.

The extent of the observed condensation is significantly larger than what can be predicted by a simple application of the Kelvin equation. A deeper investigation of the wetting properties of the new fractured surfaces is in progress, considering the effects of charge distribution, hydration forces, small amounts of soluble contaminants and the effect of local roughness.

The existence of a 100 nm liquid condensate could have a crucial impact on the stress corrosion mechanisms. The presence of a meniscus with nanometric curvature implies a significant Laplace depression inside the condensate with resulting consequences on local mechanical and chemical properties. The limited volume of the condensate makes it very susceptible to alter its chemical composition and pH due the corrosion of the glass. This effect would further be enhanced by ion exchange processes in the case of alkali-containing glasses [10].

## ACKNOWLEDGEMENTS

This work was carried out at L.C.V.N., UMR CNRS 5587, Université Montpellier 2, France. We wish to thank P. Solignac, R. Vialla, J.M. Fromental, G. Prevot, E. Alibert, F. Célarié, A. Dittmar and C. Oelgardt for their valuable assistance. We are also grateful to C. Fretigny, E. Charlaix, L. Bocquet and M. Ramonda for fruitful discussions.

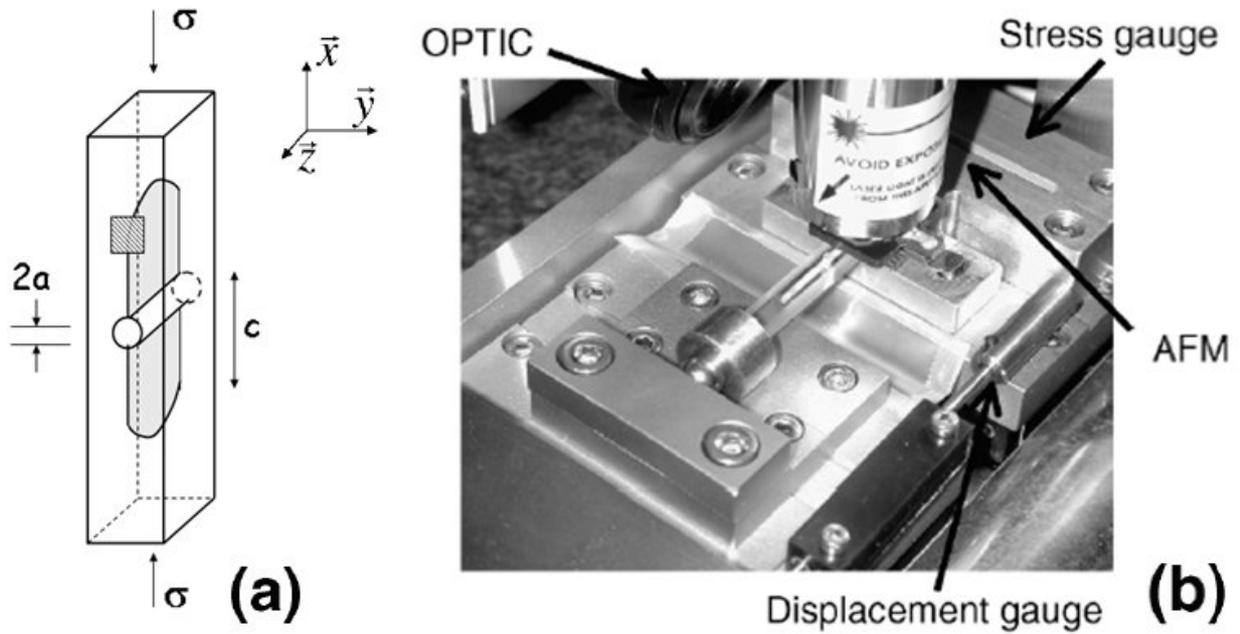

*Fig. 1 – (a) Schematic of the DCDC geometry. Two symmetric cracks (in grey) propagate in opening mode (mode I) starting from the central hole. The hatched zone corresponds to the zone which is observed by AFM as shown in the snapshot of the experiment (b).*

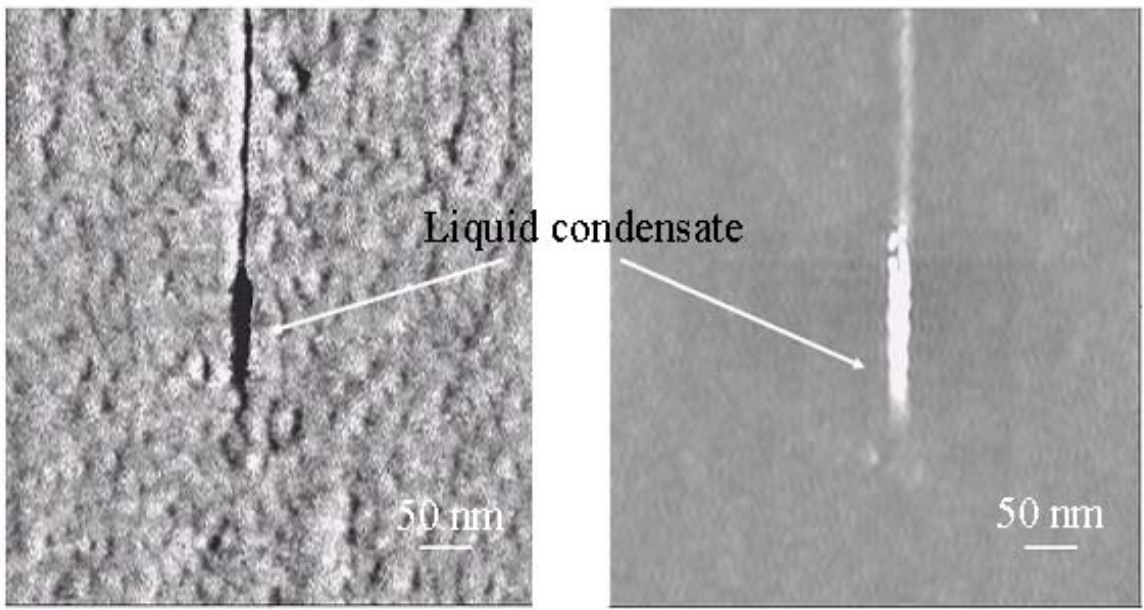

*Fig. 2 - AFM phase images of the crack tip: (left) RH = 1%, (right) RH = 70%. The phase is coded in gray levels increasing from black to white. The crack propagates from top to bottom.*

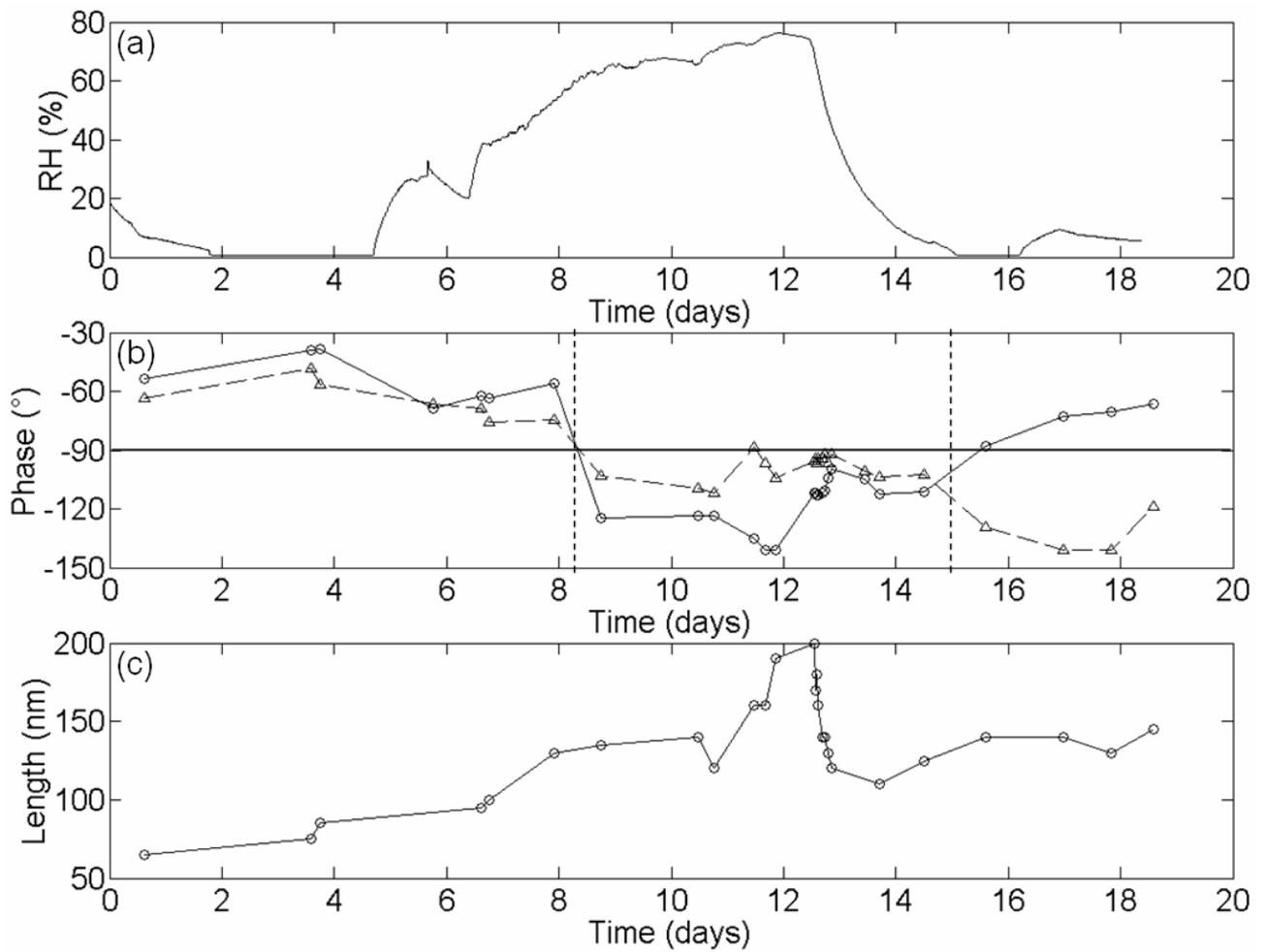

*Fig. 3 - Curves of relative humidity (a), phase (b) and condensate length (c) vs time. In graph (b), the circles (solid line) correspond to the average phase on the silica glass; the triangles (dashed line) correspond to the phase on the condensate.*